\documentclass[prl,twocolumn,fleqn]{revtex4}
\usepackage{amscd,verbatim}
\usepackage[all]{xy}
\usepackage{graphicx}

\usepackage{amsmath}
\usepackage[psamsfonts]{amssymb}
\usepackage{euscript}

\usepackage{latexsym}

\renewcommand{\(}{\begin{equation}}
\renewcommand{\)}{end{equation} \vspace{-.05in}\linebreak}

\newcounter{saveeqn}
\newcounter{savealpheqn}

\newcommand{\alpheqn}{\setcounter{saveeqn}{\value{equation}}%
  \stepcounter{saveeqn}\setcounter{equation}{0}%
  \renewcommand{\theequation}{\mbox{\arabic{section}.\arabic{saveeqn}
\alph{equation}}}
  \renewcommand{\)}{\end{equation}}}
\def\part#1{\frac{\partial}{\partial{#1}}}%
\def\group#1{\refstepcounter{equation}\setcounter{saveeqn}
 {\value{equation}}%
  \label{#1}\setcounter{equation}{0}%
\renewcommand{\theequation}{\mbox{\arabic{section}.\arabic{saveeqn}
\alph{equation}}}
  \renewcommand{\)}{\end{equation}}}
\newcommand{\reseteqn}{\setcounter{equation}{\value{saveeqn}}%
  \renewcommand{\theequation}{\arabic{section}.\arabic{equation}}%
  \renewcommand{\)}{\end{equation}}}

\newcommand{\aalpheqn}{\setcounter{saveeqn}{\value{equation}}%
  \stepcounter{saveeqn}\setcounter{equation}{0}%
  \renewcommand{\theequation}{\mbox{
        \Alph{subsection}.\arabic{saveeqn}\alph{equation}}}
   \renewcommand{\)}{\end{equation}}}
\newcommand{\areseteqn}{\setcounter{equation}{\value{saveeqn}}%
  \renewcommand{\theequation}{\Alph{subsection}.\arabic{equation}}%
  \renewcommand{\)}{\end{equation}}}

\renewcommand{\thefootnote}{\alph{footnote}}
\renewcommand{\(}{\begin{equation}}
\renewcommand{\)}{\end{equation}}
\newcommand{\ba}{\begin{eqnarray}}
\newcommand{\ea}{\end{eqnarray}}

\newcommand{\bp}{\mathop{\vtop{\ialign{##\crcr
   $\hfil\displaystyle{}\hfil$\crcr\noalign{\kern-13pt\nointerlineskip}
   \BIG{(}\hskip0pt\crcr\noalign{\kern3pt}}}}}
\newcommand{\cbp}{\mathop{\vtop{\ialign{##\crcr
   $\hfil\displaystyle{}\hfil$\crcr\noalign{\kern-13pt\nointerlineskip}
   \BIG{)}\hskip0pt\crcr\noalign{\kern3pt}}}}}
\newcommand{\pa}{\mathop{\vtop{\ialign{##\crcr
    
$\hfil\displaystyle{\oplus}\hfil$\crcr\noalign{\kern+1pt\nointerlineskip 
}
   \hspace{.08in}$^{\alpha=0}$\hskip6pt\crcr\noalign{\kern3pt}}}}}
\renewcommand{\sp}{,\hspace{.3in}}

\newcommand{\R}{\ensuremath{\mathbb R}}

\newcommand{\cpt}{\ensuremath{{\mathbb C {\text{P}}^2}}}

\newcommand{\Z}{\ensuremath{\mathbb Z}}

\newcommand{\beq}{\begin{equation}}
\newcommand{\eeq}{\end{equation}}

\numberwithin{equation}{section}
\renewcommand{\theequation}{\mbox{\arabic{equation}}}

\catcode`\@=11
\def\vereq#1#2{\lower3pt\vbox{\baselineskip1.5pt \lineskip1.5pt
\ialign{$\m@th#1\hfill##\hfil$\crcr#2\crcr\sim\crcr}}}
\catcode`\@=12

\makeatletter
\newcommand\figcaption{\def\@captype{figure}\caption}
\newcommand\tabcaption{\def\@captype{table}\caption}
\makeatother
\renewcommand{\(}{\begin{equation}}
\renewcommand{\)}{\end{equation}}

\newcommand{\CC}{{\mathbb C}}
\newcommand{\RR}{{\mathbb R}}
\newcommand{\ZZ}{{\mathbb Z}}

\newcommand{\two}{\text{I}\!\text{I}}
\newcommand{\twoa}{\text{I}\!\text{IA}}
\newcommand{\twob}{\text{I}\!\text{IB}}

\newcommand{\RP}{\RR \text{P}}
\newcommand{\rp}{\RP}

\begin{document}
\def\thefootnote{\fnsymbol{footnote}}

\title{On the Topology and H-Flux of T-Dual Manifolds}

\author{Peter Bouwknegt$^{1,2}$, Jarah Evslin$^3$ and Varghese Mathai$^2$}
\affiliation{
$^1$Department of Physics, 
University of Adelaide, Adelaide, SA 5005, Australia \\
$^2$Department of Pure Mathematics, University of Adelaide,
Adelaide, SA 5005, Australia \\
$^3$INFN Sezione di Pisa, Via Buonarroti, 2, Ed.~C, 56127 Pisa, Italy}

\begin{abstract}
\noindent
We present a general formula for the topology and $H$-flux of the T-dual  
of a type \two\ compactification.  Our results apply to T-dualities  
with respect to any free circle action.  In particular we find that the  
manifolds on each side of the duality are circle bundles whose  
curvatures are given by the integral of the dual $H$-flux over the dual  
circle.  As a corollary we conjecture an obstruction to multiple  
T-dualities, generalizing the obstruction known to exist on the twisted  
torus.  Examples include $SU(2)$ WZW models, Lens spaces and the  
supersymmetric string theory on the non-$spin$ 
$\text{AdS}^5\times \CC P^2\times  S^1$ compactification.
\end{abstract}

%
\setcounter{footnote}{0}
\renewcommand{\thefootnote}{\arabic{footnote}}


\maketitle

\renewcommand{\thepage}{\arabic{page}}


T-duality is a generalization of the $R\to 1/R$ invariance of string
theory compactified on a circle of radius $R$.  The local
transformation rules of the low energy effective fields under
T-duality, known as the Buscher rules \cite{Bus},
have been known for some time.  However in cases in  
which there is a topologically nontrivial NS 3-form $H$-flux the  
Buscher rules only make sense on each local spacetime patch.  Several  
examples of T-duals to such backgrounds have been found  
\cite{AABL,DLP,GLMW,KSTT}  and in each case it was seen that T-duality  
changes not only the $H$-flux but also the spacetime topology.  
While any new set of equivalences between compactifications may be
useful for model-building,
T-dualities involving $H$-flux have also led to the surprising discovery 
of the new physical phenomenon ``supersymmetry without supersymmetry''
in Refs.~\cite{Bak,DLP}.

In this  
letter, which is a physicists perspective of Ref.~\cite{us}, we will present a general formula for  
the topology and $H$-flux of a compactification from the topology and  
$H$-flux of its T-dual.  As supporting evidence, we have shown  
\cite{us} that locally our formula agrees with the Buscher rules and  
that globally it yields an isomorphism of the twisted K-theory-valued  
conserved RR charges \cite{MM,Wib,BM}.  Here we will demonstrate that  
our formula reproduces the old examples of topology change and easily  
generates many more.  Proofs, physical motivations, more complete  
referencing and several applications (including some tantalizingly  
mysterious clues to F-theory) may be found in the companion paper  
Ref.~\cite{us}.

Henceforth we will restrict our attention to type \twoa\ and \twob\  
string theory on 10-manifolds $E$ and $\hat{E}$, respectively, which  
admit the free circle actions
\begin{equation}
f:S^1\times E \longrightarrow E\sp
\hat{f}:\hat{S^1}\times \hat{E} \longrightarrow \hat{E}.
\end{equation}
The space of orbits of these actions are 9-manifolds which we call $M$  
and $\hat{M}$.  The freeness of the actions implies that each orbit is  
a loop and that none of these loops degenerate.  As a result $E$ and  
$\hat{E}$ are circle bundles over the bases $M$ and $\hat{M}$, and so  
their topologies are entirely determined by the topology of their bases  
together with the curvatures $F$, $\hat{F}$ (the first Chern classes of  
the bundles).  

As we will see nontrivial bundles are T-dual to configurations with
$H$-flux.  Thus we will need to include the fluxes $H$ and $\hat{H}$
in our two compactifications.  The two configurations are then
topologically determined by the triplets $(M,F,H)$ and
$(\hat{M},\hat{F},\hat{H})$ where $M$ and $\hat{M}$ are 9-manifolds,
$F$ and $\hat{F}$ are two-forms on $M$ and $\hat{M}$, and $H$ and
$\hat{H}$ are three-forms on the total spaces $E$ and $\hat{E}$.  To
capture the topology of a configuration it will suffice to consider
the fieldstrengths and Chern classes as elements of integral
cohomology.  This perspective is useful in that it automatically
identifies some gauge equivalent configurations, excludes
configurations not satisfying some equations of motion and imposes the
Dirac quantization conditions.

\noindent
{\textbf{\textit{Our Result:} The compactifications topologically  
specified by ${\mathbf{{\mathit{(M,F,H)}}}}$ and  
$(\hat{M},\hat{F},\hat{H})$ are T-dual if}}
\begin{equation}
M\cong \hat{M},\quad F=\int_{\hat{S^1}}\hat{H}, \quad \hat{F}=\int_{S^1}H.  
\label{main}
\end{equation}
This condition determines, at the level of cohomology, the curvatures
$F$ and $\hat{F}$.  However the NS fieldstrengths are  
only determined up to the addition of a three-form on the base $M$,  
because the integral of such a form over the circle fiber vanishes.  We  
further impose that the two dual three-forms on $M$ must be equal, as  
is made precise in Ref.~\cite{us}, where the duality map of the RR  
fields (viewed both as elements of cohomology and of K-theory) can also  
be found.

In the remainder of this letter we will provide examples and  
applications of our result.  When the curvatures $F$ and $\hat{F}$  
are topologically trivial (in the second cohomology of $M$) the bundles  
are trivial and so our two spacetimes are both topologically the  
trivial bundle $M\times S^1$.  Using the K\"unneth formula we may  
decompose the third cohomology of the total space $M\times S^1$
\begin{align}
&  H^3(M\times S^1) = H^3(M)\otimes H^0(S^1) \oplus\nonumber \\
& H^2(M)\otimes H^1(S^1)  
= H^3(M) \oplus H^2(M) \,,
\end{align}
and so the NS fluxes $H$ and $\hat{H}$, being elements of $H^3(M\times  
S^1)$, decompose as
$H  =\alpha+\beta \,d\theta\,,
\hat{H}  =\hat{\alpha}+\hat{\beta}\, d\theta \,,$
where $\alpha,\hat{\alpha}\in H^3(M)$, 
$\beta,\hat{\beta}\in H^2(M)$, and 
$d\theta$ is the generator of $H^1(S^1)=\ZZ$.
Integrating $H$ and $\hat{H}$ over the circle, using 
the normalization $\int d\theta = 1$, 
our result yields $\alpha = \hat{\alpha}$, $\beta=\hat{\beta}=0$.
Thus we reproduce the original examples of T-duality, in which  
spacetime is the product of a 9-manifold and a circle and the $H$-flux  
is an element of the cohomology of the 9-manifold.  As expected the  
T-dual is also a product manifold and carries the same $H$-flux.

The next most trivial case is a trivial circle bundle with $H$-flux,  
which we see from (\ref{main}) is T-dual to a nontrivial bundle without  
$H$-flux.  In this case our result was demonstrated using S-duality  
and also using the $E_8$ gauge bundle formalism in Ref.~\cite{us}.

The simplest nontrivial circle bundle is the Hopf fibration over the  
2-sphere.  This bundle is constructed by cutting the 2-sphere into a  
northern $S^2_N$ and southern hemisphere $S^2_S$, over which the circle  
is trivially fibered.  The two hemispheres are then glued together along  
the equator.  In particular, each point on the equator is specified by  
a longitude $\theta$ and the attaching map identifies the point $\phi$  
on the fibre over one hemisphere to the point $\phi+\theta$ on the  
other.  The total space of this bundle is then the three-sphere $S^3$  
and the curvature $F$ is the generator $[1]$ of $H^2(S^2)=\ZZ$.

Now we may consider a type II string theory compactification on a  
3-sphere crossed with an irrelevant 7-manifold, express the 3-sphere as  
the circle bundle above, and then T-dualize with respect to the circle  
fiber.

Let us begin with a case in which there is no $H$-flux, such as a type  
\twob\ compactification of $\text{AdS}^3\times T^4\times S^3$ supported by RR  
flux.  Applying Eqn.~(\ref{main}) with $F=[1]\in H^2(S^2)$ and $H=0$ we  
find $\hat{F}=[0]$ and $\hat{H}=[1]\in H^3(S^2\times S^1)$.  Thus our  
nontrivial bundle $S^3$ becomes the trivial bundle $S^2\times S^1$,
supported by one unit of $H$-flux.  Incidentally the construction in  
Ref.~\cite{us} tells us that if there was RR 3-form flux $G_3=[k]\in  
H^3(S^3)$ on the $S^3$ then we will find $\hat G_2=[k]\in H^2(S^2)$, 
whereas  
the $G_3$-flux on the $\text{AdS}^3$ becomes $\hat G_4$-flux on 
$\text{AdS}^3\times S^1$.   
This is in accord with the usual intuition in which T-duality toggles  
whether an RR flux extends along the circle, except that here we see  
that this intuition may apply even when the circle is nontrivially  
fibered.  One might think that this would be impossible in general  
because, for example here, there is no first cohomology class  
corresponding to the circle in the 3-sphere.  This is potentially  
problematic because, for example, an arbitrary integral Romans mass  
$\hat G_0\in H^0(S^2\times S^1)$ cannot be dual to an element $G_1\in  
H^1(S^3)$ since $H^1(S^3)=0$.  Here we are saved from  
any contradiction by the supergravity equation of motion  $\hat G_0 H=0$ which  
implies that $\hat G_0=[0]$. More generally we are saved by the quantum
corrected equations of motion, which are given by the Freed-Witten 
anomaly~\cite{FW}.

A famous application of the previous example is the T-duality of string  
theory on $\R^{8,1}\times S^1$ with an NS5-brane 
extended along the plane  
$\R^{5,1}\subset\R^{8,1}$ and localized at a point $\theta\in S^1$.   
Such an NS5-brane is linked by an $S^2\times S^1$ where $S^2\subset  
\R^{8,1}$ links the $\R^{5,1}$ plane.  Recalling that NS5-branes are  
magnetic sources for NS-flux, Gauss' law allows us to integrate
$\int_{S^2\times S^1} H = 1$
and so
\begin{equation}
H=[1]\in H^3(S^2\times S^1)=\Z.
\end{equation}
As we have seen a T-duality along this circle replaces $S^2\times S^1$  
with a 3-sphere and the $H$-flux disappears.  As the $H$-flux has  
disappeared the T-dual compactification has no NS5-brane, instead it  
has been replaced by a  
circle bundle which is nontrivially fibered over each sphere linking the  
6-submanifold where the NS5-brane was.  This submanifold is now a  
Kaluza-Klein monopole for the dual $U(1)$.  Thus we have recovered the  
familiar fact that NS5-branes are T-dual to KK-monopoles (see, e.g.,  
\cite{Ton} and references therein).  If instead of a single NS5-brane  
we had considered a stack of $k$ NS5-branes then we would have had  
$H=[k]$ and so the dual bundle would again have been nontrivial, this  
time yielding a KK-monopole charge of $k$.
More generally we may consider nontrivial circle bundles and nontrivial  
$H$-flux at the same time.  For example, 
string theory on the Lens space $S^3/\Z_j$ with  
$k$ units of $H$-flux is T-dual to string theory on $S^3/\Z_k$ with $j$  
units of $H$-flux \cite{MMS}.  

An example of T-duality that has recently received attention in the  
literature \cite{KSTT} is the duality between circle bundles over a  
2-torus $T^2$.  
In particular one may start with a 3-torus,  
which is the trivial circle bundle over a 2-torus, with $k\neq 0$ units  
of $H$-flux and then T-dualize with respect to the circle fiber.  As  
above, $F=[0]\in H^2(T^2)=\Z$ and $H=[k]\in H^3(T^3)=\Z$ determine the  
dual curvature and NS flux
\begin{align}
\hat{F} & = [k]\in H^2(T^2)=\ZZ\,, \nonumber \\
\hat{H} & = [0]\in H^3(\hat E)=\Z\,.
\end{align}
Here the dual manifold $\hat E$, commonly refered to as a twisted torus or  
nilmanifold, is the circle bundle over $T^2$ characterized entirely by  
the curvature $\hat F =[k]$.

We may also try to T-dualize a larger subtorus of the original $T^3$.   
This means that after T-dualizing with respect to the fiber circle we  
may then try to T-dualize with respect to one of the circles in the  
base.  As has been found in Ref.~\cite{KSTT}, this is impossible.  In  
particular after the first circle is T-dualized, the other two circles  
have ceased to be globally defined and so cannot be T-dualized.  Had  
they both been globally defined we could have defined a  
nowhere-vanishing section of this nontrivial circle bundle.  
Thus in  
general we see that it is impossible to T-dualize with respect to any  
3-torus supporting $H$-flux.  
In fact, the example in Ref.\ \cite{KSTT} suggests the stronger result
that one cannot T-dualize on a 2-torus, unless
$\int_{T^2} H =0$ in cohomology.

A critical check of any proposed duality is that the anomalies match on  
both sides.  Although a more general matching of a particular gravitino  
anomaly was demonstrated in Ref.~\cite{us}, here we will describe one  
family of examples which illustrate the general pattern.  Consider the  
famous type \twob\ string theory compactification on 
$\text{AdS}^5\times S^5$.  
The 5-sphere $S^5$ is a circle bundle over the projective plane \cpt\  
with a single unit of curvature
$ F=[1]\in H^2(\text{AdS}^5\times \cpt)=H^2(\cpt)=\Z$.
The second equality is a result of the fact that $\text{AdS}^5$ is  
contractible and so it does not contribute to the cohomology groups; we  
may thus freely omit it from such equations.  The third cohomology  
group of this spacetime is trivial and so the $H$-flux must be  
topologically trivial.  In addition there is a RR 5-form fieldstrength  
$G_5=[N]\in H^5(S^5)$.  There is also $G_5$-flux along the noncompact  
directions.

T-dualizing with respect to the fiber circle we find that the dual  
curvature $\hat{F}$ vanishes.
The dual manifold is thus $\text{AdS}^5\times\cpt\times S^1$.   
Eqn.~(\ref{main}) is only satisfied if the dual NS fieldstrength is
\begin{align}
\hat{H} & =[1]\in H^3(\text{AdS}^5\times\cpt\times S^1) \nonumber \\ & =
H^2(\cpt)\otimes  H^1(S^1)=\Z\otimes \Z = \Z.
\end{align}
The $G_5$-flux becomes $\hat G_4=[N]\in H^4(\cpt)=\Z$ and $\hat G_6$ 
is supported  
in the noncompact directions crossed with the circle.  In particular  
there is no $\hat G_2$-flux and therefore 
the M-theory circle is trivially fibered,  
meaning that this configuration is dual to M-theory compactified on  
$\text{AdS}^5\times\cpt\times T^2$.  Notice that $\cpt$ is not $spin$, and so  
this is an M-theory compactification on a non-$spin$ manifold.
This might seem impossible, because the low energy effective  
supergravity has a gravitino and so there must be a $spin$ structure.   
But in fact this compactification was seen to be perfectly consistent  
and even supersymmetric (although the supergravity truncation is not  
supersymmetric, as the winding modes are required to complete the  
supermultiplets) in Ref.~\cite{DLP}.  To see this, notice that the  
potential gravitino anomaly should be visible in the low energy theory,  
in which we throw away KK-modes along the torus.  This leaves an  
effective 9-dimensional theory in which the gravitino is electrically  
charged under a $U(1)$ gauge fieldstrength which is the integral of $H$  
over the dual circle.  By Eqn.~(\ref{main}) this integral is just the  
original curvature $F$, so we see that our result implies that the  
original (\twob) circle is the $U(1)$ gauge bundle under which the  
gravitino is charged.  This fact would have been obvious if we had  
dimensionally reduced the \twob\ side instead.  When $F$ is odd the  
Skyrme effect then assures that the gravitino is in fact a boson  
and so its phase is well-defined even on a spacetime which is not  
$spin$.  In our case $F=[1]$ and so the compactification is consistent.

On the other hand, in the case $F=[2]$ the gravitino still behaves as a  
fermion and so the \twoa\ configuration is inconsistent.  In this case  
the \twob\ spacetime is $\text{AdS}^5\times\rp^5$ which is an inconsistent  
background in the absence of an orientifold plane.  Our result only  
applies to oriented theories, and in particular this \twob\  
compactification is inconsistent in an oriented theory, in agreement  
with the inconsistency of the \twoa\ side of the duality.  In fact, the  
total anomalies on both sides always agree: it seems as though both  
sides are free from this anomaly when the F-theory 12-manifold is  
$spin$ \cite{us}.  This connection to F-theory is tantalizing and
mysterious.

The Ro\v cek-Verlinde \cite{RV} approach  
to T-duality should be applicable to circle bundles with nontrivial
H-flux as well, hopefully  
reproducing our result directly within the non-linear sigma model context.   
In fact such an approach has already been attempted in  
Ref.~\cite{AABL}, but this program may be significantly easier with a  
proposal already at hand.  We hope that such an approach would also shed  
light on the observed obstruction to T-dualizing 2-subtorii that
support a topologically nontrivial $H$ flux.  
A more difficult problem is to generalize our results to non-free circle actions \cite{DP}, 
this may yield a formula for the topology of mirror manifolds.

\noindent
{\bf Acknowledgements} \newline
%
We would like to thank Nick Halmagyi, Michael Schulz and Eric Sharpe
for help and crucial references.  
PB and VM are financially supported by the Australian Research Council
and JE by the INFN Sezione di Pisa.


\end{document}